\documentstyle[epsfig,11pt]{article}

\newcommand{\fig}[1]{Fig.\ref{#1}}

\newcommand{\tabs}[2]{tables \ref{#1}-\ref{#2}}
\newcommand{\eqn}[1]{Eq.(\ref{#1})}

\newcommand{\ben}{\begin{enumerate}}
\newcommand{\een}{\end{enumerate}}
\newcommand{\bit}{\begin{itemize}}
\newcommand{\eit}{\end{itemize}}
\newcommand{\bc}{\begin{center}}
\newcommand{\ec}{\end{center}}
\newcommand{\bb}{\begin{bf}}
\newcommand{\eb}{\end{bf}}
\newcommand{\bsm}{\begin{small}}
\newcommand{\esm}{\end{small}}
\newcommand{\bns}{\begin{normalsize}}
\newcommand{\ens}{\end{normalsize}}
\newcommand{\bq}{\begin{equation}}
\newcommand{\eq}{\end{equation}}
\newcommand{\bqa}{\begin{eqnarray}}
\newcommand{\eqa}{\end{eqnarray}}
\newcommand{\lb}{\linebreak}

\newcommand{\vb}{\vspace*{2cm}}

\newcommand{\nn}{\nonumber}

\def\awf{a_{W\Phi}}
\def\abf{a_{B\Phi}}
\def\aw{a_{W}}

\def\el{\ell}
\def\nl{{\bar{\nu}_\ell}}

\def\c2{\chi^2}
\def\SM{Standard Model\ }
\def\notp{p\hspace*{-5.5pt}/}

\def\bqp{\bar{q}'}

\def\demo{$\Delta\eta\mu \acute{o} \kappa \varrho \iota \tau o \varsigma$}
\def\bebar{\bar{\beta}}

\begin{document}
\input feynman
 
\pagestyle{empty}
 
\begin{flushright}
DEMO-HEP-96/02 \\
September 1996
\end{flushright}
 
\vspace*{2cm}
\bc\begin{LARGE}
{\bf
ERATO: event generator for four-fermion production at LEP2 energies and beyond
}\\ \end{LARGE}
\vspace*{2cm}
{\Large
Costas G.~Papadopoulos$^\star$
} \\[12pt]
Institute of Nuclear Physics, NRCPS \demo, 15310 Athens, Greece
\\[12pt]
\vb
ABSTRACT\\[12pt]  \ec
\begin{quote}

ERATO is a Monte Carlo event generator describing
four fermion production at LEP2 and beyond. All tree-order processes leading
to four fermions are included, taking into account all relevant QCD
contributions. QED higher order corrections are introduced
via a structure function approach, whereas weak corrections related to
fermion loops are also included. Special attention has been paid
in studying the trilinear gauge couplings, and 
all possible contributions, including the CP-violating ones, 
have been implemented. 
\end{quote}

\vspace*{\fill}
\noindent\rule[0.in]{4.5in}{.01in} \\      
\vspace{.3cm}  
$^\star$ E-mail: Costas.Papadopoulos@cern.ch 

\newpage
\pagestyle{plain}
\setcounter{page}{1}

\par
\bc {\bf PROGRAM SUMMARY}\\[18pt]\ec
{\it Title of the program:} ERATO. The name originates from 
${\cal E}\varrho\alpha\tau\acute{\omega}$, the ancient
Greek Muse of lyric poetry. \\[12pt]
{\it Program obtainable from:} Dr. Costas G. Papadopoulos, Institute of Nuclear Physics,
NRCPS `Democritos', 15310 Athens, Greece and 
using ftp://alice.nrcps.ariadne-t.gr/pub/papadopo/erato.\\[12pt]
{\it Licensing provisions:} none\\[12pt]
{\it Computer for which the program is designed and others on 
which it has been tested:}
HP, IBM, ALPHA and SUN workstations.\\[12pt]
{\it Operating system under which the program has been tested:} UNIX\\[12pt]
{\it Programming language:} FORTRAN 77\\[12pt]
{\it Keywords:} four-fermion processes, trilinear gauge couplings,
event generator\\[12pt]
{\it Nature of physical problem:} 
A very important fraction of the
events produced at high-energy $e^+e^-$ collisions corresponds to
four-fermion final states. Important physical issues, like the
measurement of the mass of the $W$ boson and the study of the
trilinear gauge couplings (TGC), are based on the analysis of 
these four-fermion
final states. Other production processes, like associated Higgs production
or R-parity violating SUSY particle production, lead also to 
the same final states. It is therefore indispensable to have a 
rather accurate description of the four-fermion production, including
tree-order signal and background contributions  as well as the leading 
part of the higher order corrections. \\[12pt]
{\it Method of solution:} The construction of an event generator
is of course the desired solution to the problem of calculating
all four-fermion processes at high energies. To this end, we have calculated all
relevant matrix elements, using a variation of the spinor technique
which is more efficient in writing and testing the corresponding 
\verb+FORTRAN+ codes. A multichannel approach
on phase space generation has been used in order to deal with
the problem of the multipeak structure of the integrated function.
Higher order corrections are systematically introduced and the new physics
effects described effectively by the trilinear gauge couplings are
considered. ERATO provides the computational framework
which enables us to study in a systematic way four-fermion physics at high-energy
$e^+e^-$ colliders.

\newpage

{\Large\bf LONG WRITE-UP} 
\\[24pt]


\section{Introduction}

LEP2 is already functioning at the highest energy ever reached for 
$e^+e^-$ colliders. Moreover, machines capable to produce even higher 
energies,  up to  
the TeV scale, like a Next Linear Collider (NLC), 
are nowadays thoroughly studied \cite{zerwas}. 
Most of the physicists working in the field think that new physics
is to be expected at these energies and a lot of work has been done
more than two decades now trying to develop theoretical frameworks
to incorporate the expected `new phenomena'. On the other hand our
current understanding of the elementary particle interactions, the
so called \SM, is not yet completely tested due to the undetected Higgs particle 
a very important ingredient of the present theoretical apparatus.
Any new dynamics in the Higgs sector of the theory will have very important
consequences on the interactions among the heavy vector bosons, $W$ and 
$Z$ \cite{goun:0,goun:gauge}.
In order to study these consequences, an accurate knowledge of the
heavy vector boson production at these energies is indispensable
and since heavy vector bosons always decay into light fermions, 
the four-fermion production is undoubtly the first most interesting 
channel(s) to be studied at these energies.
On the other hand, four-fermion will be a background channel for most
of the interesting expected `new phenomena', like SUSY particle production,
new heavy vector boson contributions, etc.

ERATO is an event generator, which enables one to have an accurate
description of the four-fermion dynamics.  In this paper we will try
to give a presentation of this program, starting with the
main theoretical features and ending with a short description of how it
works with a few illustrative examples. In section 2 a brief 
presentation of the amplitude and phase space calculations is given,
in section 3 the main features of the \verb+FORTRAN+ code are described 
and finally
in section 4 certain calculations are presented.

\section{Computational Framework}

\subsection{Helicity Amplitudes}

The first important step in order to construct an event generator is of course
the calculation of all relevant tree-order matrix elements contributing
to the processes under consideration. Our first approximation consists of
neglecting the fermion masses, for all external fermion legs: this is of course
a rather good approximation, since these terms are of the order
of $m^2_{f}/s$, where $m_f$ is the mass of the light fermion and $s$ is the
square of the center of mass energy. Nevertheless, this reasoning fails in the
case of small angle scattering, a rather special
case for which an approximation will be described in the sequel.
In ERATO we have used a special representation of the so called `spinorial
structure' of the amplitudes, which follows more closely the
structure of the Feynman graph under consideration, thus facilitating the
writing and testing of the code. More specifically we are using the so called
E-vector formalism\cite{papa:sngw,papa:4f}, which 
is based on the following representation of the fermion `current',
for massless fermions:
\bq E_\lambda^\mu(p_1,p_2)\equiv\bar{u}_\lambda (p_1)\gamma^\mu u_\lambda(p_2)
\eq
where
\bqa
E_-^0&=& \sqrt{p_1^+ p_2^+}+
\frac{(p_{1x}+i p_{1y})(p_{2x}-i p_{2y})}{\sqrt{p_1^+ p_2^+}}\nn \\
E_-^x&=& \sqrt{\frac{p_2^+}{p_1^+}}(p_{1x}+i p_{1y})
+\sqrt{\frac{p_1^+}{p_2^+}}(p_{2x}-i p_{2y})\nn \\
E_-^y&=& -i\biggl( \sqrt{\frac{p_2^+}{p_1^+}}(p_{1x}+i p_{1y})
-\sqrt{\frac{p_1^+}{p_2^+}}(p_{2x}-i p_{2y})\biggr) \nn \\
E_-^z&=& \sqrt{p_1^+ p_2^+}-
\frac{(p_{1x}+i p_{1y})(p_{2x}-i p_{2y})}{\sqrt{p_1^+ p_2^+}}
\eqa
with $p^\pm=p^0\pm p^3$.

\begin{figure}[htb]
\bc 
\fbox{ 
\bigphotons
\unitlength=0.01pt
\begin{picture}(40000,16000)

\global\Xone=5000
\global\Xtwo=2500
\global\Yone=3
\global\Ytwo=4

\pfrontx = 5000
\pfronty = 5000

\drawline\fermion[\NE\REG](\pfrontx,\pfronty)[\Xone]
\drawline\fermion[\NW\REG](\pbackx,\pbacky)[\Xone]

\drawline\photon[\E\REG](\pfrontx,\pfronty)[\Ytwo]
\global\advance \pfronty by -2000 
\put(\pfrontx,\pfronty){\makebox(\plengthx,2000){$\gamma,Z$}}

\drawline\fermion[\NE\REG](\pbackx,\pbacky)[\Xone]
\global\advance \pbackx by 500 
\put(\pbackx,\pbacky){$q$}
\drawline\fermion[\SE\REG](\pfrontx,\pfronty)[\Xone]
\global\advance \pbackx by 500 
\put(\pbackx,\pbacky){$\bar{q}'$}

\global\gaplength=250
\drawline\scalar[\E\REG](\pmidx,\pmidy)[2]
\drawline\fermion[\NE\REG](\pbackx,\pbacky)[\Xtwo]
\global\advance \pbackx by 500 
\put(\pbackx,\pbacky){$\ell$}
\drawline\fermion[\SE\REG](\pfrontx,\pfronty)[\Xtwo]
\global\advance \pbackx by 500 
\put(\pbackx,\pbacky){$\bar{\nu}_\ell$}

\pfrontx = 25000
\pfronty = 5000

\drawline\fermion[\E\REG](\pfrontx,\pfronty)[10000]
\global\advance \pbackx by 500 
\put(\pbackx,\pbacky){$\bar{\nu}_e$}
\global\gaplength=250
\drawline\scalar[\N\REG](\pmidx,\pmidy)[2]
\drawline\fermion[\N\REG](\pbackx,\pbacky)[3000]
\drawline\fermion[\E\REG](\pfrontx,\pfronty)[5000]
\global\advance \pbackx by 500 
\put(\pbackx,\pbacky){$\bar{q}'$}
\global\advance \pfronty by 3000
\drawline\fermion[\E\REG](\pfrontx,\pfronty)[5000]
\global\advance \pbackx by 500 
\put(\pbackx,\pbacky){$q$}

\drawline\photon[\N\REG](\pfrontx,\pfronty)[3]
\global\advance \pmidx by 1000 
\put(\pmidx,\pmidy){{$\gamma,Z$}}
\drawline\fermion[\W\REG](\pbackx,\pbacky)[5000]
\drawline\fermion[\E\REG](\pfrontx,\pfronty)[5000]
\global\advance \pbackx by 500 
\put(\pbackx,\pbacky){$e^-$}

\put(5000,1000){\makebox(15000,2000){(a)}}
\put(25000,1000){\makebox(15000,2000){(b)}}

\end{picture}
 }
\ec
\caption[.]{Feynman graphs contributing to
$e^+e^-\to e^-\bar{\nu}_e \; q \bqp$.}
\label{fig1}
\end{figure}
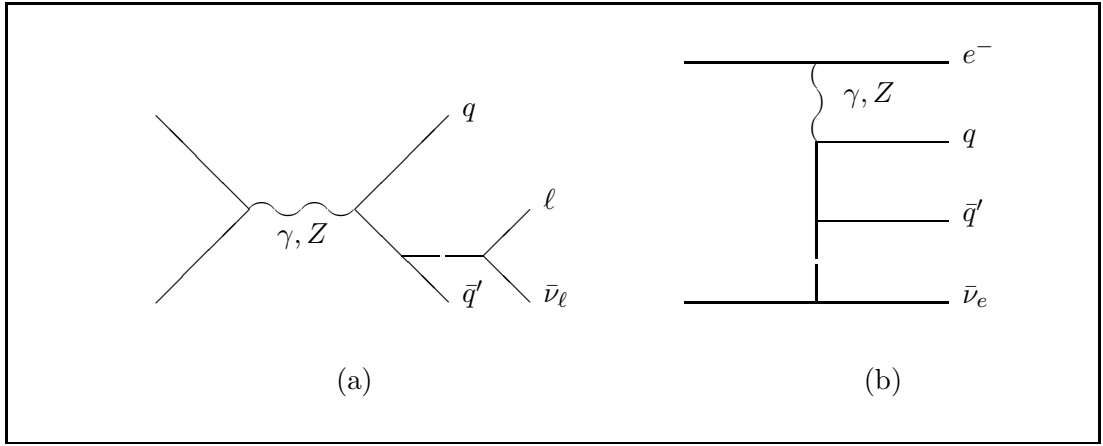

Let us see how the Feynman graph of \fig{fig1}(a), contributing to the
process $e^-(p_1)e^+(p_2)\to \ell(p_3)\bar{\nu}_\ell(p_4) q(p_5) \bqp(p_6)$,
is computed in this formulation.
First of all the spinorial part is simply written as:
\bqa &&\bar{u}_-(p_5)\gamma^\mu (-\notp_6-\notp_3-\notp_4)\gamma^\nu u_-(p_6)
\;\; \bar{u}_-(p_3)\gamma_\nu u_-(p_4) 
\;\; \bar{u}_{\lambda}(p_2)\gamma_\mu u_{\lambda}(p_1)
\nn\\
&=&-\sum_{i=3,4,6} b_i\; 
E_-(p_5,p_i)\cdot E_-(p_3,p_4) \;\; E_-(p_i,p_6)\cdot E_{\lambda}(p_2,p_1)
\eqa 
where $b_i=\pm 1$ depending on whether the particle is outgoing or incoming and 
$\lambda=\pm$ corresponds to the helicity of the incoming electron.
Then coupling constants and propagators are combined to give the final
expression. The graph of \fig{fig1}(b) is simply computed by the
interchange of $p_2 \leftrightarrow p_3$ and $b_2 \leftrightarrow b_3$.
All garphs have been computed using the above framework. Moreover where
graphs involving as external particles massless bosons, as in the
case of $e^-e^+ \to q \bar{q} gg$ which contributes to the four-jet
production channel and is experimentally indistinguishable from four-quark
production, the polarization vector of the outgoing gluon is
given by:
\bq \epsilon^\mu(p;\lambda)\equiv {1\over 2\sqrt{p\cdot q} }E^\mu_\lambda(p,q)
\eq
The auxiliary four-vector $q^\mu$ reflects the existence of the gauge
invariance, always accompanying massless spin-1 particles, and provides a very
powerful test of the computation, since any choice of $q$
($q\cdot p \neq 0$) should give {\it exactly}
the same result. Further details on the computation of the relevant 
color matrix can be found in references \cite{papa:4j,papa:2l2j}.

\subsection{Phase Space Generation}

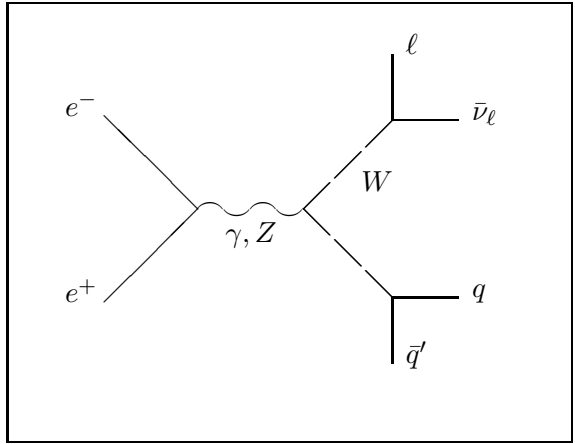
\begin{figure}[htb]
\bc 
\fbox{ 
\bigphotons
\unitlength=0.01pt
\begin{picture}(20000,16000)

\global\Xone=5000
\global\Xtwo=2500
\global\Yone=3
\global\Ytwo=4

\pfrontx = 3000
\pfronty = 5000

\drawline\fermion[\NE\REG](\pfrontx,\pfronty)[\Xone]
\global\advance \pfrontx by -1500 
\put(\pfrontx,\pfronty){$e^+$}
\drawline\fermion[\NW\REG](\pbackx,\pbacky)[\Xone]
\global\advance \pbackx by -1500 
\put(\pbackx,\pbacky){$e^-$}

\drawline\photon[\E\REG](\pfrontx,\pfronty)[\Ytwo]
\global\advance \pfronty by -2000 
\put(\pfrontx,\pfronty){\makebox(\plengthx,2000){$\gamma,Z$}}

\global\gaplength=250
\drawline\scalar[\NE\REG](\pbackx,\pbacky)[\Yone]
\global\advance \pmidx by 500 \global\advance \pmidy by -1000
\put(\pmidx,\pmidy){$W$}
\drawline\fermion[\N\REG](\pbackx,\pbacky)[\Xtwo]
\global\advance \pbackx by 500 
\put(\pbackx,\pbacky){$\ell$}
\drawline\fermion[\E\REG](\pfrontx,\pfronty)[\Xtwo]
\global\advance \pbackx by 500 
\put(\pbackx,\pbacky){$\bar{\nu}_\ell$}

\global\gaplength=250
\drawline\scalar[\SE\REG](\photonbackx,\photonbacky)[\Yone]
\drawline\fermion[\S\REG](\pbackx,\pbacky)[\Xtwo]
\global\advance \pbackx by 500 
\put(\pbackx,\pbacky){$\bar{q}'$}
\drawline\fermion[\E\REG](\pfrontx,\pfronty)[\Xtwo]
\global\advance \pbackx by 500 
\put(\pbackx,\pbacky){$q$}

\end{picture}
 }
\ec
\caption[.]{Double resonant graph contributing to
$e^+e^-\to \ell \nl \; q \bqp$.}
\label{fig2}
\end{figure}

The next important part is of course the phase space generation. This follows
closely the reasoning of references \cite{excalibur,weightopt}. The problem is that
the amplitude we have to integrate over is a very complicated
function of the kinematical variables, peaking at different regions
of phase space. The idea is to define different kinematical
mappings, corresponding to different peaking structures
of the amplitude and
then use an optimization procedure to adjust the percentage
of the generated phase-space points, according to any specific
mapping, in such a way that the total error is minimized. Let us 
present a very simple example
in order to clarify these issues. The differential cross section is written as:
\bq
d\sigma = \frac{1}{2s} |{\cal M}|^2 d\mbox{Lips}(P;p_1,\ldots,p_4)
\label{cross-section}
\eq
where 
\bq 
d\mbox{Lips}(P;p_1,\ldots,p_n)
=(2\pi)^{4-3n}\;\prod^n_{i=1}\left( d^4p_i \,\delta(p_i^2)\,\theta(p_i^0) \right) 
\delta(P-\sum^n_{i=1}p_i)\;.
\eq
For the double resonant graph of \fig{fig2} and in order to describe efficiently its
peaking structure we have naturally chosen  the variables 
$s_+\equiv p_+^2=(p_{q}+p_{\bqp})^2$ and
$s_-\equiv p_-^2=(p_{\ell}+p_{\nl})^2$ and write the invariant phase space as
\bq 
d\mbox{Lips}(P;p_{\ell},p_{\nl},p_{q},p_{\bqp})=
d\mbox{Lips}(P;p_+,p_-) \;\frac{ds_+}{2\pi}\; 
d\mbox{Lips}(p_+;p_{q},p_{\bqp})
\;\frac{ds_-}{2\pi}\; d\mbox{Lips}(p_-;p_{\ell},p_{\nl})\;.
\eq
In this way we have now the freedom to generate the integration variables $s_\pm$ such that
their distribution follows the well known Breit-Wigner form, namely 
\[
\frac{1}{\left( s_\pm-m^2_W\right)^2+m_W^2\Gamma_W^2  }\;.
\]
This is actually achieved in ERATO by the following mapping:
\bqa
m^2_{q,\bqp}&=&m_W^2+m_W\Gamma_W\tan(\rho_1 (y_+-y_-)+y_-)
\nn \\
y_+&=&\tan^{-1}\bigl(\frac{E^2-m_W^2}{m_W\Gamma_W}\bigr)
\nn \\
y_-&=&\tan^{-1}\bigl(-\frac{m_W}{\Gamma_W}\bigr)
\nn \\
m^2_{\el,\nl}&=&m_W^2+m_W\Gamma_W\tan(\rho_2 (y_+'-y_-)+y_-)
\nn \\
y_+'&=&\tan^{-1}\bigl(\frac{(E-m_{q,\bqp})^2-m_W^2}{m_W\Gamma_W}\bigr)
\nn \\
\cos\theta_W &=& 2\rho_3-1, \;\;\; \phi_W= 2 \pi \rho_4
\nn \\
\cos\theta^*_\ell &=& 2\rho_5-1, \;\;\; \phi^*_\ell= 2 \pi \rho_6
\;\;\mbox{(in the $\ell,\nl$ rest frame)}
\nn \\
\cos\theta^*_q &=& 2\rho_7-1, \;\;\; \phi^*_q= 2 \pi \rho_8
\;\;\mbox{(in the $q,\bqp$ rest frame)}
\eqa
where $\rho_i$, $i=1,\ldots,8$, represent uniformly distributed 
pseudorandom numbers in (0,1).
In order to have a rather complete description of the process
$e^+e^- \to e^-\bar{\nu}_e q \bqp$, for example, sixteen different mappings
have to be used. The number of mappings used is specific to each process
and is appropriately defined in ERATO.

The large number of mappings, whose phase-space densities are 
described by the probability distributions $g_i(\Omega)$, naturally 
introduces the following representation of the total  
phase-space density:
\bq
g(\Omega) = \sum_{i} \alpha_i\, g_i(\Omega)
\eq
where $\Omega$ stands for the phase space element and 
\bq
\int d\Omega\, g(\Omega) = \int d\Omega\, g_i(\Omega) = \sum_{i} \alpha_i = 1\;.
\eq 
The variables $\alpha_i$, are known as the {\it a priori weights}.
Their main property is that the result of the MC integration 
is independent of them:
\bq
\int d\Omega\, f(\Omega) = \left< w \right> \equiv \lim_{N\to \infty}
\frac{1}{N}\sum_{i=1}^{N} \frac{f(\Omega_i)}{g(\Omega_i)}\;. 
\eq 
On the other hand the MC error depends on $\alpha_i$: this enables us 
to perform a {\it weight optimization} which could
increase the speed of the MC integration\footnote{A detailed description can 
be found in reference \cite{weightopt}.}. This is achieved
by minimizing the variance estimator, $W(\alpha)=\left< w^2 \right>$,
with respect to $\alpha_i$, under the constraint $\sum_{i} \alpha_i = 1$.
The optimization therefore is equivalent to the evaluation of the parameter
values $\alpha^{opt}_i$,
for which the MC error is minimized. This leads to the following set 
of equations:
\[
W_i(\alpha)=W_j(\alpha)\;,\;\; \mbox{for all}\;\; i,j=1,\ldots,N
\]
where
\[
W_i(\alpha) = \left< \frac{g_i(\Omega)}{g(\Omega)} w^2 \right>
\]
with the MC weight $w$ defined by
\[
w= \frac{1}{g(\Omega)}\,\frac{1}{2s}\,|{\cal M}|^2\;. 
\]
ERATO uses an iterative algorithm to solve the above 
equations. The convergence of this procedure is checked by the
evaluation of the variable 
\bq 
{\cal D} = \max_{i,j} | W_i - W_j |
\label{distance}
\eq
which should vanish at the optimum point. 
In the numerical procedure ${\cal D}$ measures indeed how well our $\alpha$'s 
approximate the optimum solution $\alpha^{opt}_i$.

\subsection{Higher Order Corrections}
 
A substantial part of the higher order corrections to the tree-level amplitudes
have been taken into account in ERATO. As is well known the main part of these
corrections comes from the emission of soft photons from the initial
colliding particles. Although for neutral current processes, this ISR correction
is a well defined, gauge-invariant quantity, for charged currents only the
leading logarithmic (LL) part can be computed unambiguously. Nevertheless, this
LL part accounts for most of these corrections, and therefore
still it can be used to evaluate the ISR effect.  In ERATO we have implemented 
the ISR in the structure function 
approach. This simply means that the true differential cross section
is not given by \eqn{cross-section}, but by:
\bq
d\sigma_{\mbox{ISR}}= dx_1 dx_2 f(x_1) f(x_2) d\sigma(x_1 x_2 s)
\eq
where the momenta of the initial particles are given by, 
$p_{e^-}=x_1\frac{\sqrt{s}}{2}(1;0,0,1)$ and $p_{e^+}=x_2\frac{\sqrt{s}}{2}(1;0,0,-1)$.
The explicit form of the function $f(x)$ \cite{lep2ww} is:
\bqa
f(x) &=& \frac{\exp\left( \left(-\gamma_E+\frac{3}{4}\right)\beta\right)}{
\Gamma(1+\beta)} \beta (1-x)^{\beta-1} \nn \\
&-&\frac{1}{2} \beta (1+x)
-\frac{1}{8} \beta^2 \left[ \frac{1+3 x^2}{1-x} \log(x)
+4 (1+x) \log(1-x) +5 +x\right]
\label{str-fun}
\eqa
where $\gamma_E$ is the Euler constant and 
\[ \beta= \frac{\alpha}{\pi}\left( \ln\frac{s}{m_e^2} -1 \right),
\]
with $\alpha$ being the electromagnetic coupling constant.

An other important part of the higher order corrections is the so-called 
Coulomb correction. This is rather important at the threshold region,
$\sqrt{s}\sim 161$ GeV , reaching the level of 5\%. It is the
result of the long-range character of the electromagnetic interactions,
contributing a factor roughly proportional to $\alpha/\beta$, which near
threshold gives a substantial contribution. Its explicit form is

\bq   \label{CS}
  \sigma_{\mbox{Coul}} = \sigma_{\mbox{Born}}^{\tt CC3}\,
  \frac{\alpha\pi}{2\bebar}
    \left[ 1 - \frac{2}{\pi}\,\arctan\left( \frac{|\beta_M+\Delta|^2-\bebar^2}
    {2\bebar\,\mbox{Im}(\beta_M)}\right) \right],
\eq
with
\bqa \label{shorth}
  \bebar &=& \frac{1}{s}\,\sqrt{s^2-2s(p_+^2p_-^2)+(p_+^2-p_-^2)^2}
       \nn\\
  \beta_M &=& \sqrt{1-4M^2/s},
     \qquad M^2 = m_W^2 - i m_W \Gamma_W -i\epsilon  \nn\\
  \Delta &=& \frac{|p_+^2-p_-^2|}{s}  ,
\eqa
and $-\pi/2 < \arctan{y} < \pi/2$. Here $\bebar$ is the average
velocity of the $W$~bosons in their centre-of-mass system and {\tt CC3} 
refers to the three, double resonant graphs contributing to the on-shell
$W$ pair production, $e^-e^+\to W^- W^+$.

Finally an important part of the radiative corrections is related to the
width of the unstable particles, $W$ and $Z$~bosons. As it has been shown in 
references \cite{papa:4f,bhf1} the corresponding corrections, which consist
of resuming all closed fermionic-loop contributions to the two and three 
point functions, although necessary to restore
gauge invariance of the calculation, is rather small at LEP2 energies,
with the exception of final state topologies where an outgoing electron 
(or positron) is very close, essential parallel to the beam axis.

In order to describe the small angle scattering of processes with an $e^-$ or
$e^+$ in the final state we have to take into account the small electron mass,
$m_e$, leading to a further complication of the computational procedure.
Nevertheless a very good estimate of the total massive cross section can be
obtained by a rather simple approximation. This is achieved, at the leading
logarithmic level\cite{excalibur}, by introducing a cut on the
angle of the outgoing, {\it massless} electron (or positron) given by:
\bq
\theta_m= \frac{m_e \left( p^0-q^0 \right)}{p^0q^0}\;,
\label{cut-angle}
\eq
where $p^0$ ($q^0$) is the energy of the incoming (outgoing) electron.
In all cases we have checked, this approximation gives a rather good 
estimate of the total massive cross section.

We conclude this section by underlining the fact that ERATO has historically
been designed for TGC
studies and followed an older MC for single~$W$ production \cite{papa:sngw}. 
Moreover in the present version of ERATO, all TGC deviations, including CP violating 
ones, are considered\footnote{ For a detailed presentation of TGC interactions, 
see reference\cite{lep2tgc} }.

\section{Program Structure}

\begin{table}[htb]
\bc
\begin{small}
\begin{tabular}{|c|c|c|}
\hline 
  flag \verb+IPRO+ & processes & type of graphs
\\ \hline 
1 &
$\begin{array}{l}
   \nu_\mu \bar{\nu}_\mu \nu_\tau \bar{\nu}_\tau 
\end{array}$
& NC6
\\ \hline 
2 &
$\begin{array}{l}
   \nu_\mu \bar{\nu}_\mu\nu_\mu \bar{\nu}_\mu \\ 
   \nu_\tau \bar{\nu}_\tau\nu_\tau \bar{\nu}_\tau  
\end{array}$
& NC12
\\ \hline
3 &
$\begin{array}{l}
   \nu_\mu \bar{\nu}_\mu \tau^+ \tau^- \\ 
   \nu_\tau \bar{\nu}_\tau \mu^+ \mu^-  
\end{array}$
& NC10
\\ \hline
4 &
$\begin{array}{l}
   \mu^+ \mu^- \tau^+ \tau^-
\end{array}$
& NC24
\\ \hline
5 &
$\begin{array}{l}
   \mu^+ \mu^- \mu^+ \mu^- \\ 
   \tau^+ \tau^-\tau^+ \tau^-
\end{array}$
& NC48
\\ \hline
6 &
$\begin{array}{l}
   e^+ e^- \nu_\mu \bar{\nu}_\mu  \\ 
   e^+ e^- \nu_\tau \bar{\nu}_\tau  
\end{array}$
& NC20
\\ \hline
7 &
$\begin{array}{l}
   e^+ e^- \mu^+ \mu^-  \\ 
   e^+ e^- \tau^+ \tau^- 
\end{array}$
& NC48
\\ \hline
8 &
$\begin{array}{l}
   e^+ e^- e^+ e^-  
\end{array}$
& NC144
\\ \hline
\end{tabular}
\end{small}
\caption[.]{Physical processes in program llll\_n.}
\label{tab1}
\ec
\end{table}

\begin{table}[htb]
\bc
\begin{small}
\begin{tabular}{|c|c|c|}
\hline 
  flag \verb+IPRO+ & processes & type of graphs
\\ \hline 
1 &
$\begin{array}{l}
   \mu^+ \mu^-   \nu_\mu \bar{\nu}_\mu \\
   \tau^+ \tau^- \nu_\tau \bar{\nu}_\tau 
\end{array}$
& NC12+CC7
\\ \hline 
2 &
$\begin{array}{l}
   e^+ e^-   \nu_e \bar{\nu}_e 
\end{array}$
& NC24+CC14+CC18
\\ \hline
3 &
$\begin{array}{l}
   \mu^+ \mu^-   \nu_e \bar{\nu}_e  \\ 
   \tau^+ \tau^- \nu_e \bar{\nu}_e 
\end{array}$
& NC12+CC9
\\ \hline
4 &
$\begin{array}{l}
   \nu_\mu \bar{\nu}_\mu \nu_e \bar{\nu}_e \\
   \nu_\tau \bar{\nu}_\tau \nu_e \bar{\nu}_e  
\end{array}$
& NC6+CC6
\\ \hline
5 &
$\begin{array}{l}
   \nu_e \bar{\nu}_e \nu_e \bar{\nu}_e 
\end{array}$
& NC18+CC18
\\ \hline
\end{tabular}
\end{small}
\caption[.]{Physical processes in program llll\_c1.}
\label{tab2}
\ec
\end{table}

\begin{table}[htb]
\bc
\begin{small}
\begin{tabular}{|c|c|c|}
\hline 
  flag \verb+IPRO+ & processes & type of graphs
\\ \hline 
1 &
$\begin{array}{l}
   \mu^- \bar{\nu}_\mu \tau^+ \nu_\tau   \\
   \mu^+ \nu_\mu \tau^- \bar{\nu}_\tau  
\end{array}$
& CC9
\\ \hline 
2 &
$\begin{array}{l}
   e^- \bar{\nu}_e \mu^+ \nu_\mu   \\ 
   e^- \bar{\nu}_e \tau^+ \nu_\tau \\ 
   e^+ \nu_e \mu^- \bar{\nu}_\mu   \\ 
   e^+ \nu_e \tau^- \bar{\nu}_\tau \\ 
\end{array}$
& CC18
\\ \hline
\end{tabular}
\end{small}
\caption[.]{Physical processes in program llll\_c2.}
\label{tab3}
\ec
\end{table}

\begin{table}[htb]
\bc
\begin{small}
\begin{tabular}{|c|c|c|}
\hline 
  flag \verb+IPRO+ & processes & type of graphs
\\ \hline 
1 &
$\begin{array}{l}\\[-10pt]
   \nu_\mu \bar{\nu}_\mu D \bar{D}  \\
   \nu_\tau \bar{\nu}_\tau D \bar{D}   
\end{array}$
& NC10
\\ \hline 
2 &
$\begin{array}{l}\\[-10pt]
   \nu_\mu \bar{\nu}_\mu U \bar{U}  \\
   \nu_\tau \bar{\nu}_\tau U \bar{U}   
\end{array}$
& NC10
\\ \hline
3 &
$\begin{array}{l}\\[-10pt]
   \mu^+ \mu^- U \bar{U}  \\
   \tau^+ \tau^- U \bar{U}  \\
\end{array}$
& NC24
\\ \hline
4 &
$\begin{array}{l}\\[-10pt] 
   \mu^+ \mu^- D \bar{D}  \\
   \tau^+ \tau^- D \bar{D}   
\end{array}$
& NC24
\\ \hline
5 &
$\begin{array}{l}\\[-10pt] 
   e^+ e^- U \bar{U}   
\end{array}$
& NC48
\\ \hline
6 &
$\begin{array}{l}\\[-10pt]  
   e^+ e^- D \bar{D} 
\end{array}$
& NC48
\\ \hline
\end{tabular}
\end{small}
\caption[.]{Physical processes in program llqq.}
\label{tab4}
\ec
\end{table}

\begin{table}[htb]
\bc
\begin{small}
\begin{tabular}{|c|c|c|}
\hline 
  flag \verb+IPRO+ & processes & type of graphs
\\ \hline 
1 &
$\begin{array}{l}\\[-10pt]   
   \nu_e \bar{\nu}_e U \bar{U}
\end{array}$
& NC12+CC7
\\ \hline 
2 &
$\begin{array}{l}\\[-10pt]   
   \nu_e \bar{\nu}_e D \bar{D}
\end{array}$
& NC12+CC7
\\ \hline
\end{tabular}
\end{small}
\caption[.]{Physical processes in program neneqq.}
\label{tab5}
\ec
\end{table}

\begin{table}[htb]
\bc
\begin{small}
\begin{tabular}{|c|c|c|}
\hline 
  flag \verb+IPRO+ & processes & type of graphs
\\ \hline 
1 &
$\begin{array}{l}\\[-10pt]
   \mu^- \bar{\nu}_\mu U \bar{D} \\   
   \tau^- \bar{\nu}_\tau U \bar{D}   
\end{array}$
& CC10
\\ \hline 
2 &
$\begin{array}{l}\\[-10pt]
   e^- \bar{\nu}_e U \bar{D} 
\end{array}$
& CC20
\\ \hline
3 &
$\begin{array}{l}\\[-10pt]
   \mu^+ \nu_\mu U \bar{D} \\   
   \tau^+ \nu_\tau U \bar{D}   
\end{array}$
& CC10
\\ \hline
4 &
$\begin{array}{l}\\[-10pt]
   e^+ \nu_e D \bar{U} 
\end{array}$
& CC20
\\ \hline
\end{tabular}
\end{small}
\caption[.]{Physical processes in program evud.}
\label{tab6}
\ec
\end{table}

\begin{table}[htb]
\bc
\begin{small}
\begin{tabular}{|c|c|c|}
\hline 
  flag \verb+IPRO+ & processes & type of graphs
\\ \hline 
1 &
$\begin{array}{l}\\[-10pt]
   U \bar{U} D \bar{D} 
\end{array}$
& CC11+NC24+QCD8
\\ \hline 
2 &
$\begin{array}{l}\\[-10pt]
   U \bar{U}^\prime D^\prime \bar{D} 
\end{array}$
& CC11
\\ \hline
3 &
$\begin{array}{l}\\[-10pt]
   U \bar{U} U \bar{U} 
\end{array}$
& NC48+QCD16
\\ \hline
4 &
$\begin{array}{l}\\[-10pt]
   D \bar{D} D \bar{D} 
\end{array}$
& NC48+QCD16
\\ \hline
5 &
$\begin{array}{l}\\[-10pt]
   U \bar{U} U^\prime \bar{U}^\prime 
\end{array}$
& NC24+QCD8
\\ \hline
6 &
$\begin{array}{l}\\[-10pt]
   U \bar{U} D^\prime \bar{D}^\prime 
\end{array}$
& NC24+QCD8
\\ \hline
7 &
$\begin{array}{l}\\[-10pt]
   D \bar{D} D^\prime \bar{D}^\prime 
\end{array}$
& NC24+QCD8
\\ \hline
8 &
$\begin{array}{l}\\[-10pt]
   U \bar{U} g g \\
   D \bar{D} g g
\end{array}$
& QCD8
\\ \hline
\end{tabular}
\end{small}
\caption[.]{Physical processes in program qqqq.}
\label{tab7}
\ec
\end{table}

The complete file system is organized as follows: 

\ben

\item \verb+ffiles+ contains the seven \verb+FORTRAN+ files (\verb+*.f+) where
the computation of the corresponding processes is performed.
More specifically, \tabs{tab1}{tab7} give the physical 
processes which can
be obtained by each \verb+FORTRAN+ program. Also shown is the corresponding value
of the flag \verb+IPRO+ which is used as an input. When more than 
one physical processes are assigned to a given value of the \verb+IPRO+, 
it means that all these processes share identical description. Finally the
number of Feynman graphs contributing to the current procces is also
presented. The nomenclature NC24, CC10, etc stands for Neutral or Charged
Currents and is related to the unstable states contributing to a given process.

\item \verb+inputs+ contains the corresponding input files used in
the actual version. 

\item \verb+demo+ contains the corresponding \verb+Make+ file used 
for compilation and linking as well as a test run output.

\item \verb+share+ contains all commonly used subroutines and functions,
such as the function \verb+EVECTOR+ and the pseudo-random number generator(s). 
\een

The codes are very flexible and easily accessible to the user, so that
any update can easily be implemented. It should be noted however that
no special \verb+FORTRAN+-optimization has been performed in the writing of the code. 
Nevertheless
the time needed for event generation is rather small, so for all
usual applications such an optimization is not really necessary.

Let us start with the main common variables used in the program.
The first of course is \verb+P(1:4,1:20)+, where all particle momenta 
are stored: \verb+P(4,1:20)+ refers to the energy and \verb+P(1:3,1:20)+
to the three-momentum, $p_x,\,p_y,\,p_z$. At the beginning of each program 
(e.g. \verb+llll_n.f+), the momentum assignment
of the final particles is explicitly given, 
e.g. \lb $\mu^-(p_3)\bar{\nu}_\mu(p_4) u(p_9) \bar{d}(p_{10})$. 
Variable \verb+P(1:4,1)+ refers
to the incoming electron and \verb+P(1:4,2)+ to the incoming positron.
Included in the same common \verb+MOM+ is the array \verb+B(1:20)+
which as explained in the previous section, takes the values $\pm 1$
if the particle is outgoing (incoming). Physical constants are included in
the common \verb+PHYS+.

\begin{figure}[h]
\bc 
\fbox{ 
\bigphotons
\unitlength=0.01pt
\begin{picture}(40000,37500)

\global\Xone=5000
\global\Xtwo=2500
\global\Yone=3
\global\Ytwo=4

\pfrontx = 15000
\pfronty = 20000

\put(15000,20000){\framebox(10000,5000){ }}
\put(15000,22500){\makebox(10000,2500){EVENT} }
\put(15000,20000){\makebox(10000,2500){$w_i$} }
\put(20000,25000){\vector(0,1){5000}}
\put(20000,30000){\vector(0,-1){5000}}
\put(15000,30000){\framebox(10000,5000){ }}
\put(15000,32500){\makebox(10000,2500){DRIVE}}
\put(15000,30000){\makebox(10000,2500){${1\over N}\sum w_i$}}

\put(15000,22500){\vector(-1,0){2500}}
\put(12500,22500){\vector(1,0){2500}}
\put(2500,20000){\framebox(10000,5000){ }}
\put(2500,22500){\makebox(10000,2500){ISR}}
\put(2500,20000){\makebox(10000,2500){$x_1, x_2 $}}

\put(19000,15000){\vector(0,1){5000}}
\put(19000,15000){\vector(-1,0){2500}}
\put(6500,12500){\framebox(10000,5000){ }}
\put(6500,15000){\makebox(10000,2500){ADDRESS}}
\put(6500,12500){\makebox(10000,2500){ $\sum\alpha_i\; g_i(\Omega)$ }}

\put(11500,12500){\vector(0,-1){2500}}
\put(11500,10000){\vector(0,1){2500}}
\put(6500,5000){\framebox(10000,5000){ }}
\put(6500,7500){\makebox(10000,2500){ ALGOR's }}
\put(6500,5000){\makebox(10000,2500){ $\varrho_i \to \Omega$ }}

\put(21000,15000){\vector(0,1){5000}}
\put(21000,15000){\vector(1,0){2500}}
\put(23500,12500){\framebox(10000,5000){ }}
\put(23500,15000){\makebox(10000,2500){MASTER}}
\put(23500,12500){\makebox(10000,2500){$|{\cal M}|^2$}}

\put(28500,12500){\vector(0,-1){2500}}
\put(28500,10000){\vector(0,1){2500}}
\put(23500,5000){\framebox(10000,5000){ }}
\put(23500,7500){\makebox(10000,2500){ EVECTOR }}
\put(23500,5000){\makebox(10000,2500){ $E_{\lambda}(p,q)$ }}

\put(25000,22500){\vector(1,0){2500}}
\put(27500,22500){\vector(-1,0){2500}}
\put(27500,20000){\framebox(10000,5000){ }}
\put(27500,22500){\makebox(10000,2500){CCUTS}}
\put(27500,20000){\makebox(10000,2500){ {\small any cut} }}

\end{picture}
}
\ec
 \caption[.]{The flow-chart of the program.}
 \label{fig3}
 \end{figure}
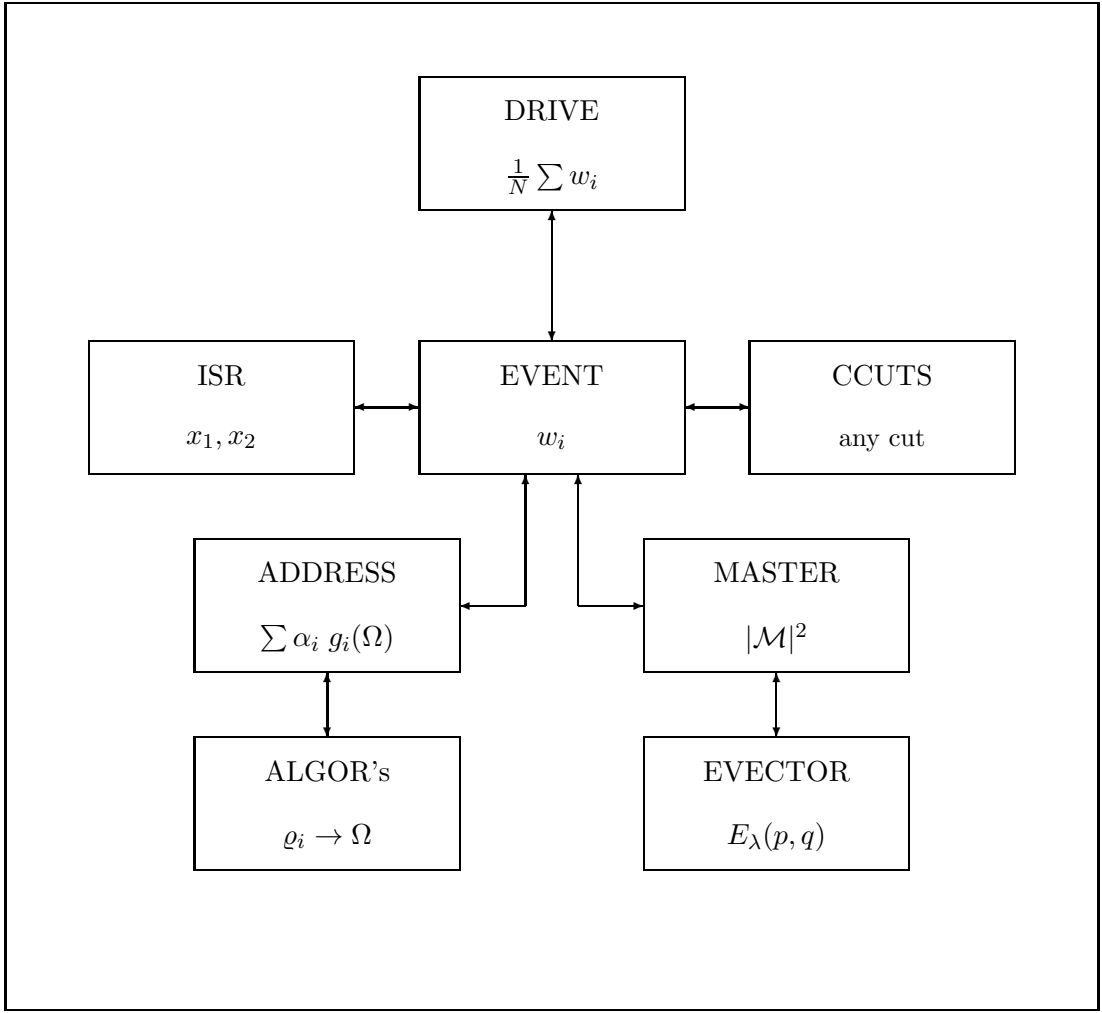

The main routine where the helicity amplitudes are calculated, is called 
\verb+MASTER+.
The array \verb+WTI(1:2,1:NITER)+ gives the amplitude squared, for 
$\mp$ helicities of the incoming electron, whereas \verb+NITER+ refers to
different input values for non-standard TGC, with \verb+ITER=1+ always
returning the \SM value.
\verb+MASTER+ is called by the routine \verb+EVENT+ after the phase-space point
generation has been performed. This is done by first producing the appropriate
energy fractions \verb+X1+ and  \verb+X2+, when ISR is on (flag \verb+ISR=1+),
which define the reduced c.m.s. energy, $s=x_1 x_2 s_0$. Notice that
the generation of these variables follows the function \verb+FISR(X)+ 
which is the product of the corresponding structure functions
given by \eqn{str-fun}. Then the routine \verb+ADDRESS+ is called, 
which
returns the phase space point, according to the given phase-space mapping, 
labeled 
by the flag \verb+IGEN+. The flag \verb+IFLAG+ is used as follows: when 
\verb+IFLAG=0+, the phase space point is generated whereas for \verb+IFLAG=1+
the phase space density is computed. Of course, as explained in the 
previous section, the distribution
among the different mappings is governed by the values of $\alpha_i$, 
\verb+ALPHA(1:NGEN)+, where \verb+NGEN+ is the total number of mappings used in
the actual calculation. Optimization on the values of $\alpha_i$, which means
redefinition of them, is performed
after certain iterations defined by  the variable \verb+IBASE+.
\verb+EVENT+ is finally returning the value of the variable 
\verb+WTI(1:2,1:NITER)+, which corresponds to the MC weight of the given
phase-space point (event).
The driving routine is called \verb+DRIVE+, which takes the weight coming from
\verb+EVENT+ and then performing all relevant sums.
A schematic representation of the flow of the program is given in \fig{fig3}.

For the cases where $WW\gamma$ and $WWZ$ vertices appear, as for example
in the \verb+evud.f+ code, the flag \verb+INAIVE+ is used to distinguish between
different ways of treating the width of the unstable particles.
As explained in detail in reference \cite{bhf1}, the fixed width solution,
corresponding to the \verb+INAIVE=1+ can be safely used by default. 
\verb+INAIVE=0+ and \verb+INAIVE=2+ correspond to the `fudge-factor' 
schemes \cite{baur+verm+zepp},
whereas \verb+INAIVE=5+ to the running width, without corrections to 
the $WW\gamma$ and $WWZ$ vertices, which of course at small momentum transfer
gives inconsistent results. Finally \verb+INAIVE=3+ correspond to the 
inclusion of all the imaginary-parts of the fermion-loop corrections, in a way
consistent with both $U(1)$ and $SU(2)$ gauge invariance.

In order to avoid singularities of the massless amplitude as well to
be as close as possible to the experimental picture cuts are applied
in the usual sense. In the present version we have implemented the
so-called Canonical Cuts, used in the LEP2 workshop analyses. Of course
any cut can be trivially implemented. In the case of ISR, there is an
additional cut on the reduced energy of the $e^+e^-$ system, defined by the
variable \verb+SCUT+.

All matrix elements have been tested against the MadGraph \cite{madgraph} calculations
to sixteen digits accuracy. Also consistency checks related to $U(1)$ and
$SU(2)$ gauge invariance, or equivalently to the {\it high-energy unitarity} 
have been performed.
Finally ERATO has been participated to the comparisons tests made at the
LEP2 workshop \cite{lep2gen}, where a very accurate check on the 
generator as a whole has been successfully performed.

\section{Test Run}

A typical input file is as follows:  
\begin{verbatim}
1                                            *process
1                                            *iterations
1                                            *ISR
0                                            *ICOULOMB
128.07 0.2310309 91.1888 2.4974 80.23 2.033  *input parameters
0                                            *cuts(total xs)
1.0 1.0                                      *cmax(cmin) cmas
175                                          *energy
200000                                       *nev
2500                                         *IEV
1                                            *0 no optimization
1                                            *naive
1.0 2. 1.5 2.0  0.8  0.5  0.0
0.0 1. 0.5 1.0 -0.2 -0.5 -1.0
0.0 1. 0.5 1.0 -0.2 -0.5 -1.0
2                                            *icase
\end{verbatim}
This is read by the \verb+evud+ program. The process under consideration
is defined by the flag
\verb+IPRO+. The \verb+NITER+ corresponds to how many iterations
the code will perform in order to calculate the weight for \verb+NITER+
different values of the TGC parameters (default=1). Then \verb+ISR+ defines
the use of initial state radiation (on=1, off=0). The same for \verb+ICOULOMB+
which is the case of the Coulomb correction.
The input parameters correspond to $1/\alpha_{em}$, 
$\sin^2\theta_w$, $m_Z$, $\Gamma_Z$, $m_W$, $\Gamma_W$ respectively.
The flag  \verb+ITOTAL+, in case it takes the value 1, will use the
approximation described by \eqn{cut-angle} for computing an estimate of the 
total massive cross section. \verb+CMAX+(\verb+CMIN+) and \verb+CMASS+ are
variables used by the generator algorithms and can be put to their limits,
\verb+CMAX=1+ and \verb+CMASS=0+ respectively. 
Then the energy and the number of MC iterations (\verb+NEV+) are given.
Variable \verb+IEV+ defines the first time when optimization will take place and the 
flag \verb+IOPT+ switch on (off) this procedure. \verb+INAIVE+ governs the
width scheme to be used, with \verb+INAIVE=1+ being the fixed width, 
and \verb+INAIVE=5+ the naive
running width. Several other options are available for specific processes, 
which can be found in the distributed codes.
Then \verb+NITER+ values for TGC parameters with \SM values 1 and 0
respectively are read by the program (the remaining terms are not read).
Finally the flag \verb+ICASE+ pick up the desired TGC parameter-relation
scheme\footnote{For a detailed analysis on these TGC parameters, see
reference \cite{papa:optobs}.}, known as $\abf$ (\verb+ICASE=0+), $\awf$ (\verb+ICASE=2+) 
and $\aw$ (\verb+ICASE=3+) as well as other choices
which can easily be found in the distributed subroutine \verb+MASTER+.

The output, apart form monitoring printings, looks as follows:

\begin{verbatim}
 ISR= 1
 ICOULOMB= 1
 STARTING AT--------Wed Jul 31 22:18:49 1996
 THIS IS THE STARTING POINT
 ITOTAL =  0
 CMAX =  1.0  CMASS =  .5
 E= 175.0
 NEV= 250000
 1./ALPHA,ZLO,ZETA,GEUL,GAB
 137.036 25.53137686693119 5.698195932780109E-02 .5772156649015319
 .9701624225753434
 E0,ELMAS,ZETA1,SCUT
 175.0 5.000000000000000E-04 17.54941409170019 1.0
 RM48 INITIALIZED:         0           0         0
 CANONICAL CUTS 1.0 3.0 .984807753012208 .9961946980917455 25.0 1

 ICASE= 2

 IMPROVMENT FOUND
 DISTANCE= 6.855963070506428E-06
 EVENTS 2500.0 2500
 IMPROVMENT FOUND
 DISTANCE= 1.797322661221867E-06
 EVENTS 7500.0 10000
 IMPROVMENT FOUND
 DISTANCE= 1.019962512017920E-06
 EVENTS 10000.0 20000
 IMPROVMENT FOUND
 DISTANCE= 7.980700337882255E-07
 EVENTS 20000.0 40000
 IMPROVMENT FOUND
 DISTANCE= 6.970462448481489E-07
 EVENTS 40000.0 80000
 IMPROVMENT FOUND
 DISTANCE= 6.011073515995931E-07
 EVENTS 80000.0 160000

 ALPHA( 1) =  .000
 ALPHA( 2) =  .298
 ALPHA( 3) =  .006
 ALPHA( 4) =  .003
 ALPHA( 5) =  .003
 ALPHA( 6) =  .006
 ALPHA( 7) =  .683

 ENERGY =   175.0000    

 ANOMALOUS MAG.MOMENT OF W=    
 .00000E+00  1.0000  1.0000  .00000E+00  .00000E+00     


   SIGMA=  .4875463E-03  +-   .1390471E-05NB


   SIGMA=  .3478089E-05  +-   .1243032E-07NB

 TOTAL SIGMA        ERROR      
 .491024E-03 .138875E-05

ICASE=   2
   [TOTAL,PASS,FAIL,ICCUT]   :  250000241945  8055  4789  1479   815   972


 GENERATOR 784 78252 3938 3030 2968 4178 156850 0 0 0 0 0 0 0 0 0 0 0 0
 0
 [MAXIMUM WT=]      1.489366913311956E-02
 ENDING AT----------Wed Jul 31 23:56:49 1996
\end{verbatim}

This is a run for the process $e^- e^+ \to \mu^- \bar{\nu}_\mu u \bar{d} $. 
The variables \verb+SIGMA+ give the left and right handed helicity
contributions with respect to the initial electron. The  
\verb+TOTAL SIGMA+ and  \verb+ERROR+ are the total cross section and its
MC error. The flag \verb+IMPROVMENT FOUND+ refers to the result of the
optimization and the final values of $\alpha_i$ are also printed out.
The variable \verb+DISTANCE+ refer to ${\cal D}$ as defined in \eqn{distance}.
Finally the values of the canonical cuts, $E_\ell$, $E_{jet}$, 
$\cos\theta_\ell$, $\cos\theta_{(\ell,jet)}$ and $m^2_{jet-jet}$ 
are presented after the flag
\verb+CANONICAL CUTS+. All the other elements in the output are
more or less self-explanatory.

\vspace*{1cm}
\noindent {\Large\bf Acknowledgements} \\[12pt]
I would like to thank the Department of Physics of the University of Durham,
where a substantial part of this work has been done. 
It is also a pleasure to thank the DELPHI colaboration and especially, 
M.~Gibbs, H.~Phillips, R.~Sekulin 
and S.~Tzamarias for their helpful suggestions and support during the writing of the
generator. I would like also to thank G.~Daskalakis for proofreading the
manuscript. This work was partially supported by the EU grant CHRX-CT93-0319.

\end{document}